\documentstyle [12pt] {report}
\begin{document}
\hfuzz=100pt
\textheight 23.0cm
\topmargin -0.5in
%
%
%
%
\newcommand{\be}{\begin{equation}}
\newcommand{\ee}{\end{equation}}
\newcommand{\bea}{\begin{eqnarray}}
\newcommand{\eea}{\end{eqnarray}}
\begin{titlepage}
\begin{flushright}
IC/92/73 \\
May, 1992
\end{flushright}
\vspace{2cm}
\begin{center}
{ \large \bf A Non Degenerate Semi-Classical Lagrangian for Dilaton-Gravity
in Two Dimensions }\\
\vspace{2cm}

{\large\bf Noureddine Mohammedi} \footnote{nour@itsictp.bitnet}\\
\vspace{.5cm}
\large International Centre for Theoretical Physics \\
P. O. Box 586, 34100 Trieste, Italy.\\

\baselineskip 18pt

\vspace{.2in}

\vspace{1cm}
{\large\bf Abstract}
\end{center}
An action for two dimensional gravity conformally coupled to two
dilaton-type fields is analysed. Classically, the theory has some
exact solutions. These include configurations representing black holes.
A semi-classical theory is obtained by assuming that these singular
solutions are caused by the collapse of some matter fields. The semi-classical
equations of motion reveal then that any generic solution must have
a flat geometry.\\

\end{titlepage}
\baselineskip 20pt

\setcounter{chapter}{1}
\setcounter{section}{1}
\setcounter{subsection}{1}
\subsection*{1.\ \,\,Introduction}
\setcounter{equation}{0}

One of the challenging issues in modern physics is the reconciliation between
quantum mechanics and general relativity. This clash followed Hawking's
discovery that matter in an initially pure quantum state can collapse to form a
black hole, which then evaporates into a mixed quantum state through Hawking
radiation [1,2,3]. Insight into this matter might be gained by analysing the
problem in the context of two dimensions. This is maily because in two
dimensions, a closed form expression for the vacuum expectation value of
the stress-energy tensor can be computed for a conformally flat space-time
metric [4]. This fact enables the calculation of the temperature and
energy flux at infinity of an evaporating black hole [5,6,7,8]. In four
dimensions, however, the conformal anomaly does not uniquely determine the
one-loop effective action, and hence the vacuum expectation value of
the energy-momentum tensor [8].
\par
An interesting model for analysing the process of black hole formation
under gravitational collapse and Hawking radiation was proposed by Callan,
Giddings, Harvey and Strominger in ref.[9]. This consists in coupling
two dimensional gravity to a dilaton field and some conformal matter fields.
At the classical level, they found a set of exact solutions among which one is
a black hole formed by  an infalling shock wave. They then, in order to
describe the back-reaction on the metric due to Hawking radiation, modified
their classical
action by including the one-loop conformal anomaly. Furthermore, by allowing
for a large number of matter fields, they argued that quantum corrections
would prevent the formation of black hole singularities in a gravitational
collapse and therefore pure quantum states would not evolve into mixed quantum
states. However, it was shown in refs.[10,11] that in this theory gravitational
collapse always ends  with the formation of a singularity. Hence the puzzles of
black hole evaporation are still persistent.\footnote{Despite this, many
interesting
papers were devoted to this model [12].}
\par
In this paper we present a model for two dimensional gravity in which
gravitational collapse does not develop a singularity. It consists of
two dimensional gravity conformally coupled to two dilaton fields.
The model is obtained by a dimensional reduction of four dimensional
Einstein gravity conformally coupled to a massless scalar field.
We begin in section two by introducing
our general model and describe some classical solutions. These include the
usual dilaton vacuum and the static black hole solutions. We then include, in
section three, the effect of some quantum matter fields and take into account
their one-loop conformal anomaly. A semi-classical treatment reveals then that
for any generic solution the metric is always flat. We also give some
exact solutions to the semi-classical equations of motion.

\setcounter{chapter}{2}
\setcounter{section}{1}
\setcounter{subsection}{1}
\subsection*{2.\ \,\,The Conformally Invariant Lagrangian}
\setcounter{equation}{0}

Recently, a model for two dimensional gravity has been put forward [9]. This is
given by the action
\be
S_{flat}=\int d^2x\sqrt{-g}\left (\omega R+\Lambda\right)\,\,\,,
\ee
where $\Lambda$ is a cosmological constant and $\omega (x)$ is a scalar field
playing the role of a dilaton field. The equations of motion resulting from
the variation of $S_{flat}$ with respect to $\omega$ and $g^{\mu\nu}$ are,
respectively, given by \footnote{Our conventions for the Riemann and Ricci
tensors are those of Misner, Thorne and Wheeler [13].}
\bea
&R=0&\nonumber\\
&g_{\mu\nu}\nabla^2\omega -\nabla_\mu\nabla_\nu\omega
-{1\over 2}\Lambda g_{\mu\nu}=0&
\,\,\,.
\eea
The first equation shows that the metric is {\it {flat}} (in two dimensions,
the vanishing of the Ricci scalar is the same as the vanishing of the
Riemann tensor). Taking the trace of the second equation and going to
light-cone coordinates, one finds the following solution for $\omega$ [14]
\bea
&\omega (x^+,x^-)={1\over 2}\left[M-\Lambda
(x^+-x^+_0)(x^--x^-_0)\right]&\nonumber\\
&\eta_{+-}=-1,\,,\,\eta_{++}=\eta_{--}=0\,\,,\,\,x^{\pm}={1\over \sqrt{2}}
(t\pm r)&\,\,\,,
\eea
with $x_0^\pm$ and $M$ being integration constants.
\par
It is therefore clear that if the "physical metric" is $g_{\mu\nu}$ then
the above model does not have a black hole solution. However, upon the
following
redefinition of the metric
\be
g_{\mu\nu}=\widetilde g_{\mu\nu}e^{-2\phi}\,\,,\,\,\omega\equiv
e^{-2\phi}\,\,\,.
\ee
The resulting action is the so-called "string-inspired" model of two
dimensional
gravity [9]
\be
S_{curved}=\int d^2x\sqrt{-\widetilde g}e^{-2\phi} \left (\widetilde R
+4\widetilde g^{\mu\nu}\partial_\mu\phi\partial_\nu\phi+\Lambda\right)\,\,\,,
\ee
where we have used the well-known result
\be
R=e^{2\phi}(\widetilde R +2\widetilde\nabla^2\phi)\,\,,\,\,
\widetilde\nabla^2={1\over\sqrt{-\widetilde g}}\partial_\mu(\sqrt{-\widetilde
g}
\widetilde g^{\mu\nu}\partial_\nu)\,\,\,.
\ee
\par
It follows that if we take $\widetilde g_{\mu\nu}=g_{\mu\nu}/\omega$ as the
"physical metric", then the model we started with exhibits a black hole
solution. It is crucial to notice that the above transformation on the metric
cannot be compensated by a variation of the field $\omega(x)$ in such a way
that the action (2.1) remains invariant. It is therefore natural to ask what
should
we take as the physical  metric of two dimensional gravity ?
\par
In this note
we present a model for two dimensional gravity which is not sensitive
to any local rescalings of the metric. That is a conformally invariant
field theory. Hence, all metrics in this theory are equivalent.
The action for such a model is given by
\bea
S_{conf}&=&\int d^2x\sqrt{-g}\left[\phi^2F(\lambda\phi)+A\lambda^2\phi^2 R+
B\phi^2\nabla_\mu\lambda\nabla^\mu\lambda\right.\nonumber\\
        &+&\left.C\phi\lambda\nabla_\mu\lambda\nabla^\mu\phi
+D\lambda^2\nabla_\mu\phi\nabla^\mu\phi\right]\,\,\,,
\eea
with $\lambda(x)$ and $\phi(x)$ being two scalar fields and $F(\lambda\phi)$
is an arbitrary function of their product. The  constant
coefficients $A$, $B$, $C$ and
$D$ are determined by  requiring conformal invariance of (2.7). Indeed, our
action
is  invariant under the following conformal transformations
\bea
g_{\mu\nu}&\rightarrow& g_{\mu\nu}e^{2\rho(x)}\nonumber\\
\lambda&\rightarrow& \lambda e^{\rho(x)}\nonumber\\
\phi &\rightarrow& \phi e^{-\rho(x)}
\eea
provided that
\bea
A&=&-{1\over 4}(C-2D)\nonumber\\
B&=&C-D\,\,\,.
\eea
The above action in (2.7) is also attractive from another point of view. In
fact, it turns out that if we choose
\be
C={4\over 3}\,\,,\,\,D=1\,\,,\,\,A={1\over 6}\,\,,\,\,B={1\over 3}\,\,,\,\,
F(\lambda\phi)={1\over 3}
\ee
then our model derives from the action functional for a real  massless
scalar field $\phi$ conformally coupled to Einstein gravity in four space-time
dimensions upon imposing spherical symmetry [15]. The four dimensional action
is written as [8]
\be
S_{(4D)}=\int d^4\hat x\sqrt{-\hat g}\left[{1\over 6}\phi^2\hat R +
\hat g^{ab}\partial_a\phi\partial_b\phi\right]\,\,\,,
\ee
where the indices $a$, $b$ range from 1 to 4 and $\hat g_{ab}$ is the four
dimensional metric. This action is invariant under the conformal
transformations
\bea
\hat g_{ab}(\hat x)&\rightarrow& \hat g_{ab}(\hat x)e^{2\rho(\hat
x)}\nonumber\\
\phi(\hat x) &\rightarrow& \phi(\hat x) e^{-\rho(\hat x)}\,\,\,.
\eea
Spherical symmetry is imposed through the ansatz:
\bea
ds^2&=&g_{\mu\nu}(x)dx^\mu dx^\nu+\lambda^2(x)S_{ij}(y)dy^idy^j\nonumber\\
\phi&=&\phi(x)\,\,\,,
\eea
where $g_{\mu\nu}$, $(\mu$,$\nu$=1,2), is the metric on a two dimensional
manifold in some coordinate patch $x^\mu$, and $S_{ij}$, $(i$,$j$=3,4), is
the metric on the standard two-sphere with coordinates $y^i$. The dimensionally
reduced action is, up to a total derivative and an overall constant volume of
the two-sphere, of the form of the model (2.7) with coefficients as given
in (2.10).
\par
In what follows and for simplicity,
we will be dealing  only with the model specified by the
coefficients written in (2.10). It is convenient to introduce the new variables
$\theta$
and $\psi$ as
\be
\phi=e^{-\theta}\,\,\,,\,\,\,\lambda=\psi e^{\theta}\,\,\,.
\ee
The action corresponding to the coefficients (2.10) is then
\be
S_{conf}={1\over 3}\int d^2x\sqrt{-g}\left[e^{-2\theta} +{1\over
2}\psi^2R+\nabla_\mu\psi\nabla^\mu\psi-2\psi\nabla_\mu\theta\nabla^\mu\psi\right]
\,\,\,.
\ee
The equations of motion are given by
\bea
{\delta S\over\delta\theta}&=&-{2\over 3}e^{-2\theta}+{2\over 3}
\nabla^\mu(\psi\nabla_\mu\psi)=0\nonumber\\
{\delta S\over\delta\psi}&=&{1\over 3}\psi R+{2\over 3}\psi
\nabla^2\theta -{2\over 3}\nabla^2\psi=0\nonumber\\
{\delta S\over\delta g^{\mu\nu}}&=&-{1\over 6}g_{\mu\nu}
\left(e^{-2\theta}-\nabla^\alpha
\psi\nabla_\alpha\psi -2\psi\nabla^\alpha\theta\nabla_{\alpha}\psi
-2\psi\nabla^2\psi\right)\nonumber\\
&-&{1\over 3}\left(\psi\nabla_\mu\theta\nabla_\nu\psi +
\psi\nabla_\nu\theta\nabla_\mu\psi+\psi\nabla_\mu\nabla_\nu\psi\right)=0\,\,\,.
\eea
Notice that the equation of motion for $\theta$ is exactly the trace of the
equation of motion for $g^{\mu\nu}$. This means that $\theta$ is not a
dynamical
field.
\par
The above equations of motion have the vacuum solution
\bea
g_{\mu\nu}&=&\eta_{\mu\nu}\nonumber\\
e^{-\theta}&=&\alpha\nonumber\\
\psi &=&\alpha r +\gamma\,\,\,,
\eea
where $\alpha$ and $\gamma$ are two constants.
A static black hole solution is immediately obtained from the dimensionally
reduced four dimensional Schwarzschild solution in the
presence of a constant scalar field. This is given by
\bea
ds^2&=& -(1-{2M\over r})dt^2 +(1-{2M\over r})^{-1}dr^2\nonumber\\
e^{-\theta}&=&\beta \nonumber\\
\psi&=&\beta r\,\,\,,
\eea
with $\beta$ and $M$ being  constants.
\par
In order to describe the metric and the fields $\theta$ and $\psi$ due to
an infalling shell of matter fields we need to patch in a continuous manner
the vacuum solution together with the black hole one across some light-like
line. For this, we introduce the null coordinates $(u,v)$ in which the metric
takes the form
\be
ds^2=-2\left(1-{2M\over r}\right)dudv\,\,\,,
\ee
where $r$ is now a function of $u$ and $v$ and is determined by the equation
\be
{1\over\sqrt{2}}\left(v-u\right)=r+2M\ln\,\left(\vert {r\over 2M}-1\vert
\right)\,\,\,.
\ee
The patching is then carried out across the wordline of the shell
$v=v_0$, where $v_0$ is a constant. However, since equation (2.20)
has no solutions in closed form, we were not able to explicitly perform
the patching and hence find the matter fields which form the
black hole. Nevertheless, we assume in the next section that the black
hole is obtained by the collapse of some conformal matter fields.
\par
We can also look for solutions in which the metric is conformally flat
\be
g_{\mu\nu}=\eta_{\mu\nu}e^{2\sigma}\,\,\,.
\ee
In this gauge the equations of motion get modified according to
\bea
{\delta S\over \delta\theta}&=&-{2\over3}e^{-2(\theta-\sigma)}
-{2\over 3}\partial_+\partial_-\psi^2=0\nonumber\\
{\delta S\over \delta\sigma}&=&{2\over 3}e^{-2(\theta-\sigma)}
+{2\over 3}\partial_+\partial_-\psi^2=0\nonumber\\
{\delta S\over \delta\psi}&=&{4\over 3}\psi\partial_+\partial_-\sigma
-{4\over 3}\psi\partial_+\partial_-\theta+{4\over 3}
\partial_+\partial_-\psi=0\,\,\,.
\eea
There are, in addition, two constraints corresponding to the gauge fixing of
the metric, $g_{++}=g_{--}=0$,
\bea
T_{++}&=&{2\over 3}\psi\partial_+\psi\partial_+\sigma
-{2\over 3}\psi\partial_+\psi\partial_+\theta-{1\over  3}\psi
\partial_+\partial_+\psi=0
\nonumber\\
T_{--}&=&{2\over 3}\psi\partial_-\psi\partial_-\sigma
-{2\over 3}\psi\partial_-\psi\partial_-\theta-{1\over 3}\psi
\partial_-\partial_-\psi=0
\eea
Notice that the equations of motion for $\theta$ and $\sigma$ are identical and
only the combination
\be
\rho(x)=\sigma(x)-\theta(x)
\ee
appears in the above equations.
This is due to the fact that our action (2.15) is conformally invariant.
\par
A general free field solution for the equations of motion and the constraints
is found to be given by
\bea
\psi (x^+,x^-)&=& \psi_+ (x^+)-\psi_- (x^-)\nonumber\\
e^{2\rho(x^+,x^-)}&=&2{d\psi_+\over dx^+}{d\psi_-\over dx^-}\,\,\,,
\eea
where $\psi_+$ and $\psi_-$ are two arbitrary functions. In this solution,
however, the conformal factor $\sigma$ is not determined and can be any
expression. Therefore the curvature can be singular. It is clear that
this family of solutions does not include the black hole solution found
in (2.18).
\par
In the particular case when $\sigma(x)=\theta(x)$, the dilaton field $\psi$
is linear and is uniquely determined to be
\be
\psi(x^+,x^-)=ax^+-{1\over 2a}x^-+b\,\,\,,
\ee
where $a$ and $b$ are constants of integration.
This case, however, is equivalent to getting rid of the $\theta$-dependence
in the action (2.15) by redefining $g_{\mu\nu}$. Indeed, by writing
\be
g_{\mu\nu}\rightarrow \bar g_{\mu\nu}e^{2\theta}\,\,\,,
\ee
the action (2.15) reduces to
\be
S={1\over 3}\int d^2x\sqrt{-\bar g}\left[1 +{1\over
2}\psi^2\bar R+\bar g^{\mu\nu}\bar\nabla_\mu\psi
\bar\nabla_\nu\psi\right] \,\,\,.
\ee
However, since the action (2.15) is conformally invariant one can easily
prevent
the above redefinition by shifting at the same time $\theta$ and $\psi$
according to (2.14) and (2.8). As
explained earlier, this is not the case in the string inspired model.
\par

\setcounter{chapter}{3}
\setcounter{section}{1}
\setcounter{subsection}{1}
\subsection*{3.\ \,\,Including the Back-Reaction}
\setcounter{equation}{0}
So far we have treated the theory at the classical level only and without any
matter fields. A semi-classical
effective action might be obtained by including the effect of some quantum
matter fields.  In this section we are assuming that some of the black hole
solutions, especially the one in (2.18), are caused by the collapse of some
conformal matter fields.
It was shown in [9] that, when there is a number of matter
fields, the back-reaction on the geometry due to Hawking radiation  can
be accounted for by adding the one-loop anomaly term to the classical action.
In
our case and in the conformal gauge,  including  the effect of the trace
anomaly yields the following effective semi-classical action
\bea
S&=&\int d^2x\sqrt{-\eta}\left[{1\over 3}e^{2\sigma}\phi^2
+{1\over 6}\lambda^2\phi^2R
+{2\over 3}\phi^2\lambda
\eta^{\mu\nu}\partial_\mu\lambda\partial_\nu\sigma
+{2\over 3}\phi\lambda^2
\eta^{\mu\nu}\partial_\mu\phi\partial_\nu\sigma \right.\nonumber\\
&+&{1\over 3}\phi^2
\eta^{\mu\nu}\partial_\mu\lambda\partial_\nu\lambda
+{4\over 3}\phi\lambda
\eta^{\mu\nu}\partial_\mu\phi\partial_\nu\lambda
+\lambda^2
\eta^{\mu\nu}\partial_\mu\phi\partial_\nu\phi\nonumber\\
&-&\left.{1\over 2}\sum^{N}_{i=1}\eta^{\mu\nu}\partial_\mu f^i\partial_\nu f^i
+\kappa \eta^{\mu\nu}\partial_\mu \sigma\partial_\nu \sigma\right]\,\,\,,
\eea
where $f^i$ are $N$ infalling matter fields  assumed to form the classical
black hole
and the term proportional to
$\kappa\equiv {\hbar N\over 24}$ represents the back-reaction on the metric
due to the collapsing $f^i$ fields.\footnote{We are assuming that the Liouville
cosmological constant can be set to zero.}
\par
Our effective Lagrangian can be regarded as a non-linear sigma model on a three
dimensional
target space (the directions along the $f^i$ fields are flat). The metric on
this target space is determined by the kinetic term for the fields $\lambda$,
$\phi$ and $\sigma$. This is given by
\be
G_{rs}=\left(\begin{array}{ccc}
{1\over 3}\phi^2&{2\over 3}\lambda\phi&{1\over 3}\lambda\phi^2\\
{2\over 3}\lambda\phi&\lambda^2&{1\over 3}\lambda^2\phi\\
{1\over 3}\lambda\phi^2&{1\over 3}\lambda^2\phi&\kappa
\end{array}\right)\,\,\,,
\ee
The determinant of this metric is
\be
det\,(G_{rs})=-{1\over 9}\kappa\lambda^2\phi^2=-{1\over 9}\kappa\psi^2\,\,\,.
\ee
Therefore, there is no degeneration of the target space metric and the kinetic
term can be always inverted. This allows, in particular, for  the
carrying out  of a weak
field  perturbation theory in the amplitude of the field $f^i$ in
order to solve, for example, the full equations when $\kappa\neq 0$. This
is in complete contrast to the string-inspired model written in (2.5). There,
in the conformal gauge, the target space metric has a determinant of
$4e^{-2\phi}(\kappa-e^{-2\phi})$ which is degenerate when the dilaton takes
the value $e^{-2\phi}=\kappa$. This fact was responsible for the singular
solutions of the equations of motion of the string-inspired model [10,11].
\par
Now, using the variables $\theta$ and $\psi$, the  semi-classical
equations of motion are given by
\bea
{\delta S\over \delta\theta}&=&-{2\over3}e^{-2(\theta-\sigma)}
-{2\over 3}\partial_+\partial_-\psi^2=0\nonumber\\
{\delta S\over \delta\sigma}&=&{2\over 3}e^{-2(\theta-\sigma)}
+{2\over 3}\partial_+\partial_-\psi^2+4\kappa\partial_+\partial_-\sigma=0
\nonumber\\
{\delta S\over \delta\psi}&=&{4\over 3}\psi\partial_+\partial_-\sigma
-{4\over 3}\psi\partial_+\partial_-\theta+{4\over 3}
\partial_+\partial_-\psi=0
\nonumber\\
{\delta S\over \delta f^i}&=&\partial_+\partial_-f^i=0\,\,\,.
\eea
The modified constraints are
\bea
T_{++}&=&{2\over 3}\psi\partial_+\psi\partial_+\sigma
-{2\over 3}\psi\partial_+\psi\partial_+\theta -{1\over 3}\psi
\partial_+\partial_+\psi
\nonumber\\
&+&\kappa\left(\partial_+\sigma\partial_+\sigma +t_+(x^+)\right)
-{1\over 2}\sum^N_{i=1}\partial_+f^i\partial_+f^i=0
\nonumber\\
T_{--}&=&{2\over 3}\psi\partial_-\psi\partial_-\sigma
-{2\over 3}\psi\partial_-\psi\partial_-\theta -{1\over 3}\psi
\partial_-\partial_-\psi
\nonumber\\
&+&\kappa\left(\partial_-\sigma\partial_-\sigma +t_-(x^-)\right)
-{1\over 2}\sum^N_{i=1}\partial_-f^i\partial_-f^i=0\,\,\,.
\eea
The functions $t_+(x^+)$ and $t_-(x^-)$ are due to the non-locality of
the anomaly term added to the classical action and are fixed by the
asymptotic physical boundary conditions [9].
\par
The first thing to notice  here is that the equations of motion for $\theta$
and $\sigma$ imply that
\be
\kappa\partial_+\partial_-\sigma=0\,\,\,.
\ee
Therefore, the Ricci scalar
\be
R=4e^{-2\sigma}\partial_+\partial_-\sigma
\ee
must vanish when $\kappa\neq 0$.
Hence, any generic solution in the conformal gauge is flat.
Consequently, perturbing the vacuum solution (2.17) by some infalling matter
fields does not produce any singularities.
\par
The solutions for the $\sigma$ and $f^i$ fields are straightforward to find
\bea
\sigma(x^+,x^-)&=&\sigma_+(x^+)+\sigma_-(x^-)\nonumber\\
f^i(x^+,x^-)&=&f^i_+(x^+)+f^i_-(x^-)\,\,\,.
\eea
A free field solution for the fields $\psi$ and $\theta$ is
given by
\bea
\psi(x^+,x^-)&=&h_+(x^+)-h_-(x^-)\nonumber\\
\theta(x^+,x^-)&=&\sigma_+(x^+)+\sigma_-(x^-)
-{1\over 2}\ln\,\left(2{dh_+\over dx^+}{dh_-\over dx^-}\right)\,\,\,.
\eea
The constraints then lead to
\bea
\sigma_+(x^+)&=&\pm{1\over\sqrt{\kappa}}\int^{x^+}dy^+\left({1\over
2}\sum^N_{i=1}\left[
{df^i_+(
y^+)\over dy^+}\right]^2-\kappa t_+(y^+)\right)^{{1\over 2}}\nonumber\\
\sigma_-(x^-)&=&\pm{1\over\sqrt\kappa}\int^{x^-}dy^-\left({1\over
2}\sum^N_{i=1}\left[
{df^i_-(
y^-)\over dy^-}\right]^2-\kappa t_-(y^-)\right)^{{1\over 2}}\,\,\,.
\eea
\par
In conlusion we have presented a model for two dimensional gravity
conformally coupled to two dilaton fields. At the classical level we found
solutions corresponding to black holes. However, we were not able to prove
that these singular solutions are formed by gravitational collapse of a shell
of matter fields. We assumed, nevertheless, that such matter fields do exist
and constructed a semi-classical theory. We then showed that all the singular
configurations of the classical theory are absent in this semi-classical
theory.

\vspace{0.5cm}

{\bf Acknowledgements} : I would like to thank F. D. Mazzitelli for  many
useful discussions. The
financial support from IAEA and UNESCO is also hereby acknowledged.

\end{document}